**Title:**

Refractive index measurement of pharmaceutical powders in the short-wave infrared range using index matching assisted with phase imaging

Cory Juntunen[1], Adam J. Rish[2], Carl Anderson[2], Yongjin Sung[1,†]

[†] To whom correspondence should be addressed. E-mail: ysung4@uwm.edu.

1. College of Engineering and Applied Science, University of Wisconsin, Milwaukee, WI 53211, USA.

2. School of Pharmacy and Graduate School of Pharmaceutical Sciences, Duquesne University, Pittsburgh, PA 15282, USA.

**Introduction**

Refractive index, a fundamental optical property, has been utilized in pharmaceutical applications for the purposes of material characterization, formulation development, and quality assessment [1]. For example, the bulk density of pharmaceutical excipient powders at different grades were previously assessed utilizing refractive index [2]. The refractive index is also required as an input to the light scattering measurement to determine the size distribution of powder [3] and diffuse reflectance spectroscopy to measure the absorption and scattering coefficients of powder products (e.g., pharmaceutical powders and tablets) [4]. In the visible, ultraviolet, and infrared ranges, the refractive index can be measured by a variety of methods. For liquids, Abbe refractometry is usually considered as the gold standard method [5]. Using Abbe refractometry in combination with bandpass filters, the refractive index can be determined over a broad wavelength range [6]. For a planar sheet comprising multiple layers of different materials, ellipsometry can accurately provide the refractive index and thickness of each constituent layer [7]. For homogeneous microspheres, the refractive index can be measured at a high throughput by using light scattering measurements performed at multiple angular positions [8]. Phase imaging, which measures a 2D phase distribution of the light altered by the sample, can determine the refractive index of homogeneous microspheres at high accuracy, albeit at a low throughput [9, 10]. For microparticles with a complex shape or a heterogeneous refractive index distribution inside, tomographic phase imaging is necessary (e.g., [11, 12]), which combines phase imaging with computed tomography. Although tomographic phase imaging can provide the 3D refractive index distribution, the refractive index accuracy is often compromised due to the data collection for a limited angular range or a misregistration of the projection images.

In this work, we demonstrate a method to accurately measure the refractive index of complex microparticles of homogeneous composition but with arbitrary shape. This is typical of powders used in the pharmaceutical industry. The proposed method combines index matching with phase imaging. Index matching uses a series of measurements with the immersion media of different refractive indices. As the refractive index of the immersion medium approaches the sample's refractive index, the reflectance from an interface is minimized. Index matching can be applied to the particles with complex, arbitrary shapes, and the method is robust, as it does not rely on any assumptions [13]. However, using an ocular estimation of the match, the uncertainty of the refractive index measurement using index matching can be as large as 0.005 [14]. In this study, we demonstrate phase imaging can significantly improve the estimation of the match producing an order-of-magnitude smaller uncertainty. Using the index matching assisted with phase imaging, we determine the refractive index of four pharmaceutical powders (lactose monohydrate, Avicel PH-102, acetaminophen, and hydroxypropyl methylcellulose) in the short-wave infrared (SWIR) range of 1100–1650 nm.

**Materials and Methods**

Pharmaceutical powders were used as received after capturing the sieve fraction between a #325 sieve (45 µm mesh) and a #500 sieve (25 µm mesh). The powders investigated were lactose monohydrate (Foremost Farms, Rothschild, WI, USA), microcrystalline cellulose (Avicel PH102, FMC BioPolymer, Mechanicsburg, PA, USA), acetaminophen (Mallinckrodt Pharmaceuticals Inc., Raleigh, NC, USA), and hydroxypropyl methylcellulose (Pharmacoat 606, Shin-Etsu Chemical Co., Ltd., Tokyo, Japan). For comparison with literature data, we measured the refractive index of polystyrene powder (Nanochemazone, 9003-53-6).

Starting with eleven standard refractive index liquids (1.450, 1.4587, 1.490, 1.516, 1.530, 1.550, 1.570, 1.580, 1.600, 1.650, and 1.700 standardized at 589.3 nm) from Cargille Laboratories (Cedar Grove, NJ, USA), we produced refractive index liquids at finer steps in the range 1.450–1.700. In particular, to produce a liquid with the refractive index value of $n$, we mixed two liquids of refractive indices $n_1$ and $n_2$ ($n_1 < n < n_2$) at the volume fractions $f_1$ and $f_2$ determined by Eqs. (1) and (2), respectively, as suggested by the manufacturer. The mixture of oils in a centrifuge tube was homogenized using a vortex mixer.

$$f_1 = (n_2 - n)/(n_2 - n_1) \quad \text{Eq. (1)}$$

$$f_2 = (n - n_1)/(n_2 - n_1) \quad \text{Eq. (2)}$$

To determine the index match, we imaged the sample immersed in a refractive index liquid SWIR digital holographic microscopy (DHM). For the imaging, samples were prepared as follows. Two round, No. 1 coverslips with a diameter of 35 mm were cleaned using an ethanol-water (70:30, v/v) mixture and completely dried in an air-ventilated hood. A small amount of pharmaceutical powder was spread across the entire area of a round cover slip. We pipetted 100µL of refractive index oil onto the powder, then placed another coverslip on top, sandwiching the sample immersed in the refractive index oil. The samples were imaged right after the preparation. All measurements were done at room temperature (23 °C).

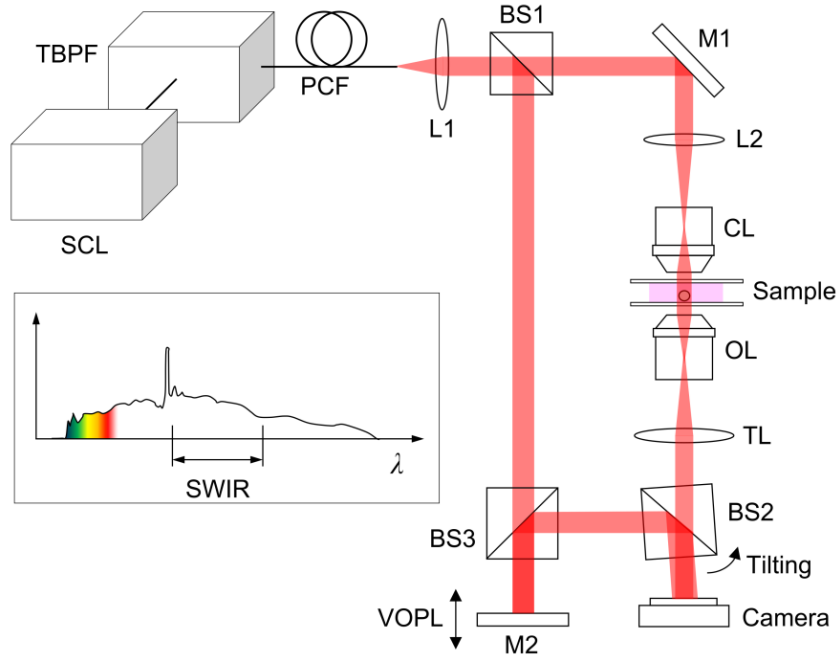

**Figure 1. Schematic diagram of the SWIR DHM system.** SCL: supercontinuum laser; TBPF: tunable bandpass filter; PCF: photonic crystal fiber; L1 and L2: achromatic lenses; M1 and M2: mirrors; BS1, BS2 and BS3: beam splitters; CL: condenser lens; OL: objective lens; TL: tube lens; VOPL: variable optical path length. A color figure is available online.

Figure 1 shows a schematic diagram of the SWIR DHM system used for hyperspectral phase imaging in the wavelength range of 1100–1650 nm. The system is built upon a Mach-Zehnder interferometer, which is modified to allow for imaging both the amplitude and phase alterations induced by the sample. For the light source, a supercontinuum laser (SCL) (NKT Photonics, WL SC400-4) is used in tandem with a tunable bandpass filter (TBPF) (NKT Photonics, SWIR HP8), which produces quasi-monochromatic light in the SWIR range at a narrow spectral band of less than 5 nm (full width at half maximum). The filtered light is coupled to the photonic crystal fiber (PCF) by using a fiber-coupling unit (NKT Photonics, SuperK Connect). The light from the PCF is collimated by the lens L1 and reaches the beam splitter BS1, where the beam is split into a sample beam path and reference beam path. The sample beam, passing through the BS1, is reflected off of the mirror M1 and illuminates the sample after being demagnified by the lens L2 and the condenser lens CL. The sample beam, whose amplitude and phase are altered by the sample, is magnified by the objective lens OL, then collimated by the tube lens TL, forming an image at the back focal plane of the TL. For both the OL and the TL, we use the same type of SWIR objectives (Edmund, 56-982) with the magnification factor of 50 and the numerical aperture of 0.65. The sample beam is combined with the reference beam using the beam splitter BS2. After being reflected off of the BS1, the reference beam passes through the beam splitter BS3 and is incident onto the mirror M2 perpendicular to the surface. The reference beam reflected off the M2 is reflected by the BS3 to the BS2, then into the camera. The InGaAs camera (Raptor Photonics, OWL 1280) has 1280×1024 pixels, each with the width of 10 µm. For off-axis digital holography [15, 16], the BS2 is slightly tilted so that the reference beam is incident onto the camera at an angle with respect to the sample beam, producing straight interference fringes on the camera. The period of the interference fringes is given by $\lambda/\sin(\theta)$, where $\lambda$ is the wavelength and $\theta$ is the angle between the sample and reference beams. The sample with a nonhomogeneous refractive index distribution alters the phase distribution within the sample beam, while the absorption by the sample alters the amplitude distribution. The straight interference fringes at the camera plane are distorted according to the altered phase distribution in the sample beam.

Noteworthy, to produce interference fringes at high visibility (or contrast), the optical path length (OPL) of the sample beam should match with that of the reference beam. This is not trivial in hyperspectral DHM because the sample and reference beams propagate through different optical elements that add OPLs varying with the wavelength. To match the OPLs of the sample and reference beams across the entire wavelength range, we mount the M2 on a motorized stage (Thorlabs, MTS50-Z8) and adjust its position as the wavelength is scanned. The optimal mirror position to produce the highest fringe contrast for each wavelength is manually determined in a calibration experiment. During the index matching experiment, a lab-made LabVIEW (National Instruments) code automatically moves the mirror M2 to the optimal mirror position for each wavelength, while scanning the wavelength and triggering the camera. The overall magnification of the SWIR DHM system is 41.7, the camera pixel resolution 0.24 µm, and the field of view 307×246 µm$^2$. Using off-axis digital holography, three camera pixels are used to sample one period of the fringe pattern; thus, the pixel resolution decreases by three times. Considering the Nyquist criterion [17], each projection image is recorded at a resolution of about 1.4 µm—six times the pixel resolution.

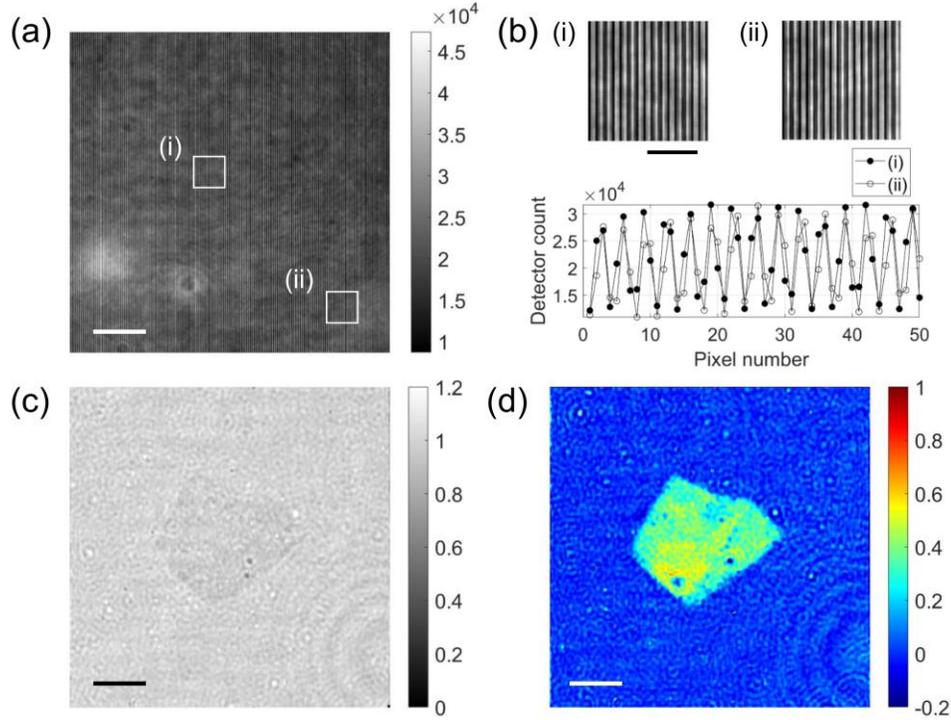

**Figure 2. Data processing for DHM.** (a) An example of raw interferogram image recorded with a lactose monohydrate powder particle at the wavelength of 1350 nm. (b) Magnified images of the regions (i) and (ii) shown in (a), and the detector counts across the interference fringes for the two cases. (c) and (d) show the amplitude and phase images, respectively, obtained from the raw interferogram shown in (a).

Figure 2(a) shows an example of the interferogram recorded with the SWIR DHM system for a lactose monohydrate powder particle at the wavelength of 1350 nm. Two square regions in the sample (i) and the background (ii) are magnified and shown in Figure 2(b). The interference fringes can be clearly seen in both the magnified images. Figure 2(b) also shows the detector counts, averaged along the fringe direction, as a function of the pixel number in the direction perpendicular to the fringes. The similar average detector counts confirm negligible absorption by the imaged powder sample, while the similar fringe visibilities indicate that the scattering by the sample is negligible. Figures 2(c) and 2(d) show the amplitude and phase images, respectively, obtained by applying a standard fringe analysis technique for off-axis digital holography [15, 16] to the recorded interferogram in Figure 2(a). The amplitude image in Figure 2(c) further confirms negligible absorption by the imaged powder sample. The phase image in Figure 2(d) can be related to the refractive index distribution within the sample. Suppose that a homogeneous specimen of the refractive index $n_s$ is immersed in a medium of the refractive index $n_m$. Assuming rectilinear propagation of the light across the sample, the sample-induced phase $\Phi(x, y)$ at the location $(x, y)$ can be related to the refractive indices $n_s$ and $n_m$ by Eq. (3).

$$\Phi(x,y) = (2\pi/\lambda)T(x,y)(n_s - n_m), \quad \text{Eq. (3)}$$

where $\lambda$ is the vacuum wavelength, and $T(x,y)$ is the sample thickness at $(x,y)$. Note that we use only the sign of $\Phi(x,y)$ in this work.

**Results and Discussion**

For each powder type, we determine the refractive index through an iterative process using the immersion media of different refractive index values. Each iteration comprises three steps: (i) perform hyperspectral phase imaging using the SWIR DHM system, then, for each wavelength, (ii) determine the sign of the phase value within the sample region and (iii) adjust the upper and lower bounds of the refractive index for the sample. If the sample produces a positive phase value, the refractive index of the current immersion medium is set as the lower bound of the sample's refractive index. If the phase value is negative, the refractive index of the current immersion medium is set as the upper bound of the sample's refractive index. In the next iteration, a new immersion medium with a different refractive index is used to reduce the gap between the upper and lower bounds. The iteration continues until the refractive index is determined within a target error, e.g., 0.001. In each iteration step, the sample's refractive index is estimated as the average of the upper and lower bounds. Noteworthy, the proposed method provides the maximum absolute error, which is given as half the difference between the upper and lower bounds, for each iteration step. This advantage is similar to that of the bracketing method (also called the bisection method) for finding the roots of an equation [18]. Other refractive index measurement methods do not provide the absolute error, as the true value is not available. Using only the sign of the sample-induced phase, the proposed method is not affected by permeation of the immersion medium into the sample. For example, a sample with a higher refractive index in vacuum than the medium would exhibit a decreased refractive index after the permeation, but it would still be higher than the medium's refractive index.

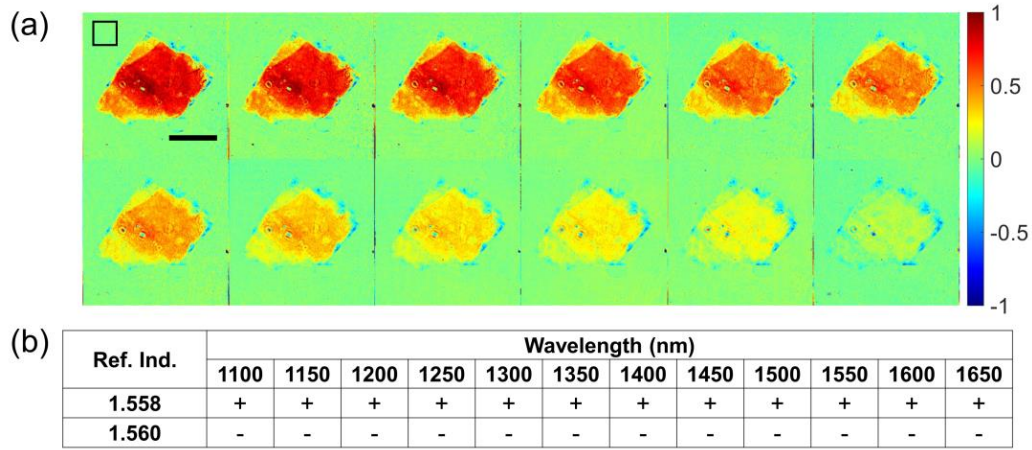

**Figure 3. Example of refractive index measurement using index matching assisted with phase imaging.** (a) An example of the phase images acquired in the 1100–1650 nm wavelength range at 50 nm steps (from top left to bottom right). The sample is lactose monohydrate in an immersion oil with the refractive index 1.558 (standardized at 589.3 nm). (b) An example of the index matching table summarizing the upper and lower bounds of the sample's refractive index at each wavelength. The plus sign indicates that the sample's refractive index is higher than that of the immersion medium, and the minus sign indicates the opposite.

Figure 3(a) provides an example of phase images obtained using lactose monohydrate in immersion oil with a refractive index of 1.558 (standardized at 589.3 nm), in the wavelength range of 1100–1650 nm at 50 nm increments. For every wavelength examined, the refractive index of the sample is higher than that of the immersion medium because the phase values in the sample region are unmistakably positive in every image. Performing the same measurement on a different sample immersed in a liquid with a refractive index of 1.560 (standardized at 589.3 nm), we obtain negative phase values in the sample region for all the wavelengths. Figure 3(b) shows an example of the index matching table for the upper and lower bounds of the refractive index for different wavelengths.

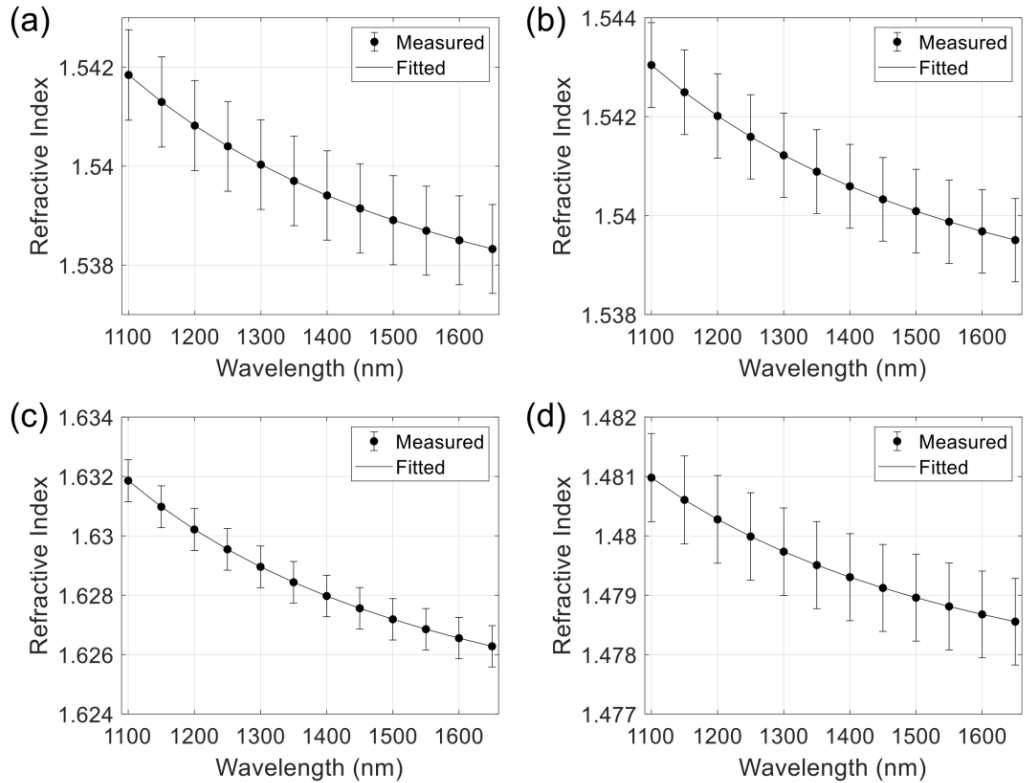

**Figure 4. Refractive index dispersion of pharmaceutical powders measured in the SWIR range of 1100–1650 nm: (a) lactose monohydrate, (b) Avicel PH102, (c) acetaminophen, and (d) hydroxypropyl methylcellulose.** The error bar represents the upper and lower bounds of the measurement at each wavelength. The solid line represents the fitting curve using Cauchy's equation with three terms.

Figure 4 shows the refractive index dispersion curves for four types of pharmaceutical powders (lactose monohydrate, Avicel PH102, acetaminophen, and hydroxypropyl methylcellulose) measured with the proposed method in the SWIR range of 1100–1650 nm. For each data point, the caps of the error bar indicate the upper and lower bounds determined with the index matching method, and the filled circle represents their average. For the measured pharmaceutical powders, each of the upper and lower bounds has been identified by the same liquid across the entire wavelength range, whose refractive index dispersion is given as Cauchy's equation. Thus, the data points marked by the filled circles (i.e., the average of the upper and lower bounds) can be exactly fit to the Cauchy's equation with three terms: $n(\lambda) = a + b/\lambda^2 + c/\lambda^4$, where $\lambda$ is the wavelength in µm, and $a$, $b$, and $c$ are fitting parameters. The fitted curve is shown as a solid line in each graph. Table I summarizes the fitting result.

Figure 5 shows the refractive index dispersion curve for polystyrene measured in the present study, together with the data available in the literature. The refractive index of polystyrene has been measured using various techniques in the visible and SWIR ranges, for example, Nikolov and Ivanov [19], Ma et al. [20], Seet et al. [21], Sultanova et al. [22], Zhang et al. [23], and Nyakuchena et al. [10]. We have extrapolated the dispersion relation to the SWIR range for comparison when the data is only available in the visible range. Our result is closest to the result obtained by Zhang et al. who measured the refractive index dispersion of polystyrene films using ellipsometry over the wavelength range of 0.4–2 µm [23]. Interestingly, large deviations are observed among the reported refractive index curves, which may be attributed to different crosslinking conditions of the polystyrene polymer, different characterization methods, or both.

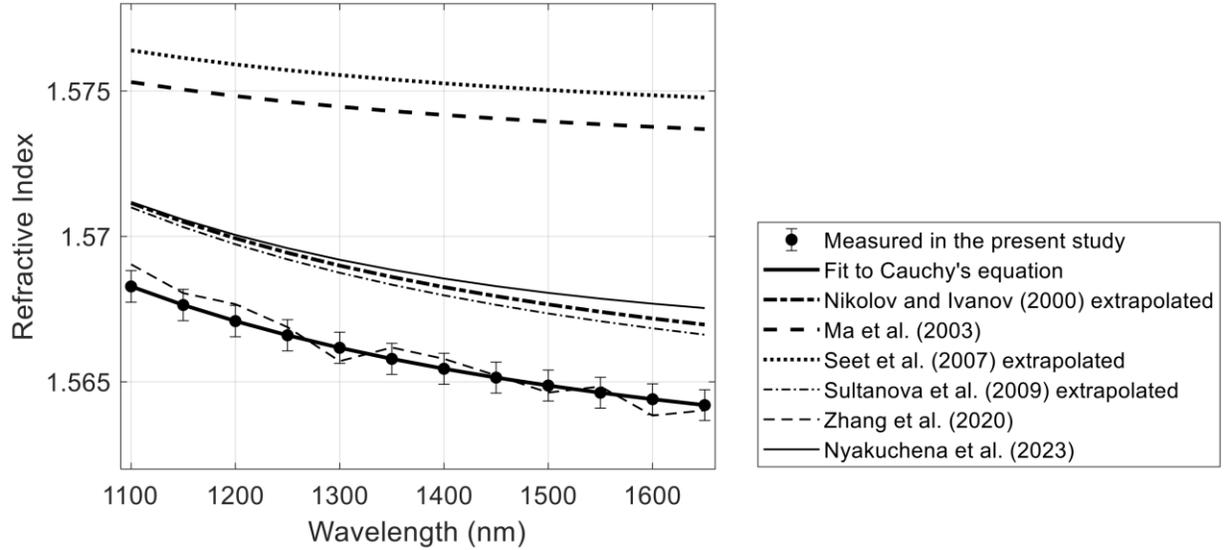

**Figure 5. Refractive index dispersion of polystyrene in the SWIR range of 1100−1650 nm.** The data measured in the present study is shown with the data available in the literature.

Table 1. Summary of the refractive index dispersion (1100–1650 nm) of
the powders determined in this study. $\lambda$ is the wavelength in µm.

| Material | Dispersion equation |
| --- | --- |
| Lactose Monohydrate | $1.536 + 7.330 \times 10^{-3}/\lambda^2 + 2.771 \times 10^{-4}/\lambda^4$ |
| Avicel PH102 | $1.537 + 7.373 \times 10^{-3}/\lambda^2 + 2.803 \times 10^{-4}/\lambda^4$ |
| Acetaminophen | $1.622 + 1.102 \times 10^{-2}/\lambda^2 + 9.520 \times 10^{-4}/\lambda^4$ |
| Hydroxypropyl methylcellulose | $1.477 + 5.140 \times 10^{-3}/\lambda^2 + 1.188 \times 10^{-4}/\lambda^4$ |
| Polystyrene | $1.561 + 8.372 \times 10^{-3}/\lambda^2 + 4.312 \times 10^{-4}/\lambda^4$ |

When phase imaging is used for a more complex task such as feature extraction, a high signal-to-noise ratio (SNR) would be needed. As we use phase imaging only to determine whether the sample's refractive index is higher or lower than that of the immersion medium, the detection limit may be defined as a signal that yields an SNR of two. Then, the precision of the refractive index measurement, or the smallest refractive index difference $\Delta n_{min}$ measurable with the proposed method may be obtained from Eq. (4).

$$\Delta n_{min} = (\lambda/\pi)\, \Delta\Phi_{noise}/T_{max}, \qquad \text{Eq. (4)}$$

where $\Delta\Phi_{noise}$ is the noise level in the phase measurement, and $T_{max}$ is the maximum thickness of a sample that can be imaged with phase imaging. The standard deviation for the square region in the upper left corner of Figure 3(a) is approximately 0.040 rad, when averaged across the 12 images, which may be considered as $\Delta\Phi_{noise}$. The decrease in fringe visibility caused by the sample's absorption and scattering characteristics determines $T_{max}$ for phase imaging based on interferometers. The intensity of the sample beam diminishes with increasing sample thickness because of absorption by the sample, and its spatial coherence decreases because of internal scattering. The interference fringes become less visible due to both the intensity reduction and the coherence loss [24, 25], which ultimately causes the phase unwrapping to fail [26].

We can estimate the thickness of the sample we imaged and use the maximum value as $T_{max}$, even though $T_{max}$ would vary depending on the type of sample. For the phase images shown in Fig. 3(a), $T_{max}$ is estimated to be about 97±36 µm. For the calculation, we used Eq. (3) together with the refractive index of the immersion medium ($n_m$), which is given by the manufacturer, and the sample's refractive index ($n_s$), which we have determined using the index matching. With $T_{max}$ set at 97 µm, $\Delta\Phi_{noise}$ of 0.040 rad translates to $\Delta n_{min}$ of $1.4 \times 10^{-4}$ at 1100 nm and $2.2 \times 10^{-4}$ at 1650 nm, which can be considered as the uncertainty in the refractive index measurement for the proposed method. The large standard deviation in $T_{max}$ is due to the small difference (0.0013) between $n_s$ and $n_m$. We note that the sample thickness can be measured at a higher precision using an immersion medium producing a larger difference between $n_s$ and $n_m$.

**Conclusion**

In this work, we demonstrated an index matching method assisted with phase imaging to measure the refractive index of pharmaceutical powders in the SWIR range of 1100–1650 nm within 0.001 accuracy. Unlike traditional approaches, which only provide the precision, the bracketing method used in this study provides the absolute error. By varying the immersion media in succession, refractive index of the sample can be determined to four decimal places. The refractive index accuracy of the proposed method will be ultimately limited by the maximum sample thickness that can be imaged with phase imaging, which is determined by the absorption and scattering properties of the sample.

**Acknowledgments**

This research was funded by the National Science Foundation (1808331).

**Additional Information**

The authors declare no competing interest.

# References

1. Mohan S, Kato E, Drennen III JK, Anderson CA (2019) Refractive index measurement of pharmaceutical solids: a review of measurement methods and pharmaceutical applications. *Journal of Pharmaceutical Sciences*, 108(11):3478–3495.

2. Stranzinger S, Faulhammer E, Li J, Dong R, Khinast JG, Zeitler JA, Markl D (2019) Measuring bulk density variations in a moving powder bed via terahertz in-line sensing. *Powder Technology*, 344:152–160.

3. Xu R (2015) Light scattering: A review of particle characterization applications. *Particuology*, 18:11–21.

4. Shi Z, Anderson CA (2010) Application of Monte Carlo simulation-based photon migration for enhanced understanding of near-infrared (NIR) diffuse reflectance. Part I: Depth of penetration in pharmaceutical materials. *Journal of pharmaceutical sciences*, 99(5):2399–2412.

5. Rheims J, Köser J, Wriedt T (1997) Refractive-index measurements in the near-IR using an Abbe refractometer. *Measurement Science and Technology*, 8(6):601.

6. Kedenburg S, Vieweg M, Gissibl T, Giessen H (2012) Linear refractive index and absorption measurements of nonlinear optical liquids in the visible and near-infrared spectral region. *Optical Materials Express*, 2(11):1588–1611.

7. Vedam K (1998) Spectroscopic ellipsometry: a historical overview. *Thin solid films*, 313:1–9.

8. Dick WD, Ziemann PJ, McMurry PH (2007) Multiangle light-scattering measurements of refractive index of submicron atmospheric particles. *Aerosol science and technology*, 41(5):549–569.

9. Fu D, Choi W, Sung Y, Oh S, Yaqoob Z, Park Y, Dasari RR, Feld MS (2009) Ultraviolet refractometry using field-based light scattering spectroscopy. *Optics express*, 17(21):18878–18886.

10. Nyakuchena M, Juntunen C, Shea P, Sung Y (2023) Refractive index dispersion measurement in the short-wave infrared range using synthetic phase microscopy. *Physical Chemistry Chemical Physics*, 25(34):23141–23149.

11. Choi W, Fang-Yen C, Badizadegan K, Oh S, Lue N, Dasari RR, Feld MS (2007) Tomographic phase microscopy. *Nature methods*, 4(9):717–719.

12. Charrière F, Marian A, Montfort F, Kuehn J, Colomb T, Cuche E, Marquet P, Depeursinge C (2006) Cell refractive index tomography by digital holographic microscopy. *Optics letters*, 31(2):178–180.

13. Meeten GH (2017) Refractive index measurement. *Measurement, Instrumentation, and Sensors Handbook*, :50–1.

14. Meeten G (1986) Refraction and extinction of polymers. *Elsevier Applied Science Publishers Ltd., Optical Properties of Polymers,* :1–62.

15. Creath K (1988) Phase-measurement Interferometry. *Progress in Optics*, :349–393.

16. Ikeda T, Popescu G, Dasari RR, Feld MS (2005) Hilbert phase microscopy for investigating fast dynamics in transparent systems. *Optics letters*, 30(10):1165–1167.

17. Gonzales RC, Wintz P (1987) Digital Image Processing.

18. Chapra SC, Canale RP (2011) Numerical Methods for Engineers.

19. Nikolov ID, Ivanov CD (2000) Optical plastic refractive measurements in the visible and the near-infrared regions. *Applied Optics*, 39(13):2067–2070.

20. Ma X, Lu JQ, Brock RS, Jacobs KM, Yang P, Hu X-H (2003) Determination of complex refractive index of polystyrene microspheres from 370 to 1610 nm. *Physics in Medicine & Biology*, 48(24):4165.

21. Seet KY, Vogel R, Nieminen TA, Knöner G, Rubinsztein-Dunlop H, Trau M, Zvyagin AV (2007) Refractometry of organosilica microspheres. *Applied Optics*, 46(9):1554–1561.

22. Sultanova N, Kasarova S, Nikolov I (2009) Dispersion properties of optical polymers. *Acta Physica Polonica A*, 116(4):585–587.

23. Zhang X, Qiu J, Li X, Zhao J, Liu L (2020) Complex refractive indices measurements of polymers in visible and near-infrared bands. *Applied optics*, 59(8):2337–2344.

24. Yashiro W, Terui Y, Kawabata K, Momose A (2010) On the origin of visibility contrast in x-ray Talbot interferometry. *Optics express*, 18(16):16890–16901.

25. Lynch SK, Pai V, Auxier J, Stein AF, Bennett EE, Kemble CK, Xiao X, Lee W-K, Morgan NY, Wen HH (2011) Interpretation of dark-field contrast and particle-size selectivity in grating interferometers. *Applied optics*, 50(22):4310–4319.

26. Judge TR, Bryanston-Cross P (1994) A review of phase unwrapping techniques in fringe analysis. *Optics and Lasers in Engineering*, 21(4):199–239.